\documentclass[preprint,journal]{vgtc}            

\onlineid{1190}

\ieeedoi{10.1109/TVCG.2024.3456196}

\vgtccategory{Research}

\vgtcpapertype{application/design study}

\title{Cultural Reflections in Virtual Reality: The Effects of User Ethnicity in Avatar Matching Experiences on Sense of Embodiment}

\author{%
  \authororcid{Tiffany D.\ Do}{0000-0003-3323-4586},
  \authororcid{Juanita Benjamin}{0000-0003-1308-6678},
  Camille Isabella Protko, and 
  \authororcid{Ryan P.\ McMahan}{0000-0001-9357-9696}
}

\authorfooter{
  \item
  	Tiffany D. Do is with Drexel University.
  	E-mail: tiffany.do@drexel.edu.
  \item
  	Juanita Benjamin is with The University of Central Florida.
  	E-mail: juanita.benjamin@ucf.edu.
  \item
  	Camille Isabella Protko is with The University of Central Florida.
  	E-mail: ca916041@ucf.edu.
  \item Ryan P. McMahan is with Virginia Tech.
  	E-mail: rpm@vt.edu.
}

\abstract{%
  Matching avatar characteristics to a user can impact sense of embodiment (SoE) in VR. However, few studies have examined how participant demographics may interact with these matching effects. We recruited a diverse and racially balanced sample of 78 participants to investigate the differences among participant groups when embodying both demographically matched and unmatched avatars.  We found that participant ethnicity emerged as a significant factor, with Asian and Black participants reporting lower total SoE compared to Hispanic participants. Furthermore, we found that user ethnicity significantly influences ownership (a subscale of SoE), with Asian and Black participants exhibiting stronger effects of matched avatar ethnicity compared to White participants. Additionally, Hispanic participants showed no significant differences, suggesting complex dynamics in ethnic-racial identity. Our results also reveal significant main effects of matched avatar ethnicity and gender on SoE, indicating the importance of considering these factors in VR experiences. These findings contribute valuable insights into understanding the complex dynamics shaping VR experiences across different demographic groups.
}


\teaser{
  \centering
  \includegraphics[width=\linewidth]{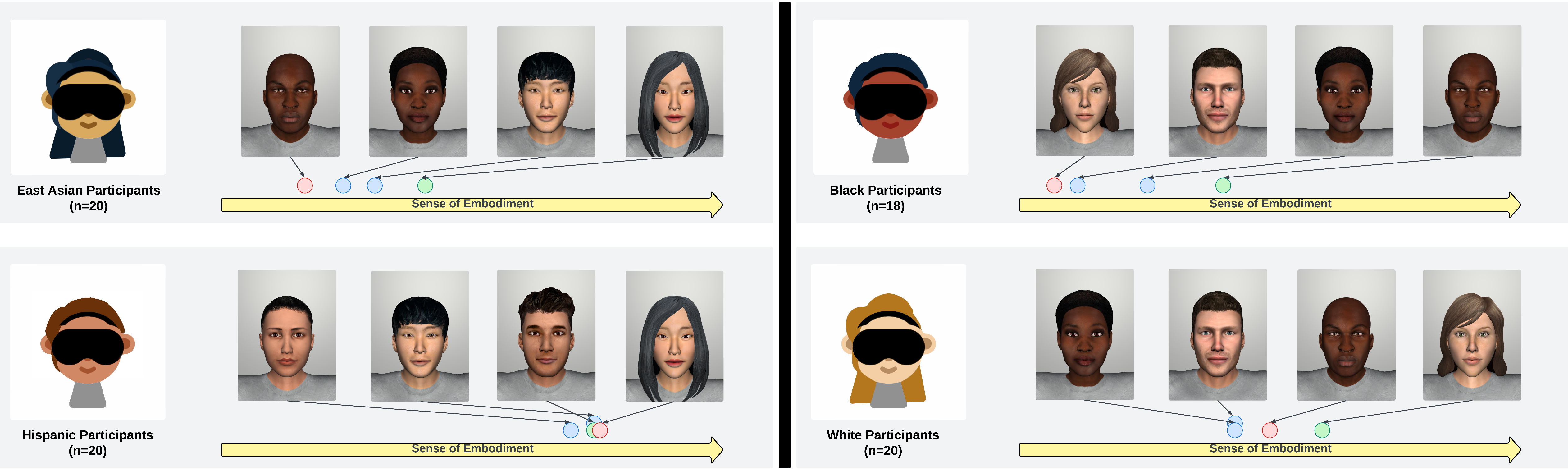}
  \caption{%
  	This figure shows how user ethnicity influences self-avatar matching effects and sense of embodiment (SoE). The green circle represents an avatar that has both ethnicity and gender matched, while the red circle represents an avatar that has neither. Blue circles represent avatars that have only one factor matched to the user. Asian and Black participants generally had lower SoE and had higher SoE with avatars that matched their ethnicity. 
  }
  \label{fig:teaser}
}

\graphicspath{{figs/}{figures/}{pictures/}{images/}{./}} 

\usepackage{tabu}                      
\usepackage{booktabs}                  
\usepackage{lipsum}                    
\usepackage{mwe}                       

\usepackage{mathptmx}                  

\usepackage{appendix}
\usepackage{amssymb}
\usepackage[normalem]{ulem}
\useunder{\uline}{\ul}{}
\usepackage{cancel}

\newcommand{\rem}[1]{}
\newcommand{\soutpars}[1]{\let\helpcmd\sout\parhelp#1\par\relax\relax}
\usepackage{booktabs}

\begin{document}


\firstsection{Introduction}

\maketitle
Recent advances in virtual reality (VR) hardware and affordability have fueled a rapid expansion of the market globally \cite{Dluhopolskyi2021}. As the demographic of VR users evolves, it is essential to understand how diverse users engage with and perceive the technology, identifying variations among user groups. However, scholars have noted that this area of work remains fairly limited. In particular, Peck et al. \cite{peck2021divrsify} highlighted the potential role of race\footnote{We use the term race to refer to a ``visually distinct social group with a common ethnicity" \cite{Rhodes2010}.} and ethnicity in VR assessments and made a call for action for more in-depth studies of these factors. 

An insightful interview study by Freeman and Maloney \cite{freeman_body_2021} found that non-White users may attribute greater significance to their self-avatar's skin tone and ethnic facial features. Despite these findings, there exists few empirical assessments comparing how diverse racial or ethnic demographics may differ in VR experiences. Moreover, limited studies have directly compared multiple racial groups of participants due to low sample sizes or a lack of diversified recruitment.

The issue of diversity extends to research regarding virtual avatars, which play an influential role in VR applications due to their prevalence in social immersive environments \cite{gonzalez-franco2020b}. In a recent study, Do et al. \cite{do2024stepping} highlighted a critical concern in virtual embodiment research: a considerable number of VR studies and popular consumer applications do not permit users to customize their self-avatar's ethnicity, which are often depicted as White, regardless of the user's ethnicity. Moreover, their findings revealed that not matching a user's ethnicity and their embodied avatar's ethnicity results in a lower sense of embodiment (SoE), highlighting the potential influence of ethnicity mismatch in VR applications and research.   For example, SoE specifically can influence perceptions of distance \cite{gonzalez-franco2019}, enhance cognitive tasks like object size estimation \cite{Jung2018}, and closely tie into one's sense of spatial presence within a virtual environment \cite{eubanks_effects_2020}.

In this work, we refer to SoE as "the ensemble of sensations that arise in conjunction with being inside, having, and controlling a body especially in relation to virtual reality applications" \cite{kilteni_sense_2012}. Moreover, avatar embodiment is defined as the "condition of owning a virtual body through
consistent sensory information" \cite{Mottelson2023} where "some properties are processed in the same way as the properties of one’s body" \cite{deVignemont2011}. 

It is important to note that Do et al. \cite{do2024stepping} did not analyze differences between different demographic groups due to a relatively low sample size, potentially limiting insights into the results. For example, embodying a different-ethnicity or different-gender avatar may elicit varied responses depending on the user's ethnicity or gender. Thus, the actual impact of constraining users to characters with predefined avatars remains unclear, as this effect may vary based on the user's characteristics. To address this gap, our study aims to replicate and extend their work by recruiting a larger and more diverse group of participants, enabling a nuanced analysis of the impact of user ethnicity and gender on self-avatar matching effects.

Our study contributes to a more thorough understanding of how user characteristics interacts with avatar matching effects and how diverse demographic groups may vary in VR experiences. We have recruited a large participant group (n=78) and ensured a balanced representation of four ethnicities and two genders, enabling detailed analyses of participant demographics. The targeted ethnicities included East or Southeast Asian (Asian), Black or African American (Black), Hispanic or Latino (Hispanic), and White participants. Our exploration is centered on unraveling how participant ethnicity and gender interact with self-avatar ethnicity-matching and gender-matching effects. To guide our investigation, we formulated the following research questions:

\begin{itemize}
    \item \textbf{RQ1:} How does user ethnicity interact with avatar matching effects, and what variations exist among different ethnic groups? 
    
    \item \textbf{RQ2:} How does user gender interact with these matching effects, and what variations exist among men and women? 
\end{itemize}

\begin{table*}[] \centering
\caption{An overview of VR studies examining demographic disparities among participants. Studies categorized by the demographic aspect under analysis (e.g., Gender, Ethnicity).}
\begin{tabular}{@{}llll|l@{}}
\hline
 & \multicolumn{3}{c|}{\textbf{Participant Groups}} & \multicolumn{1}{c}{\textbf{Measures}} \\ \hline
\multicolumn{1}{c}{} & \textbf{Gender} & \textbf{Ethnicity} & \textbf{Other} & \textbf{} \\ \hline
Freeman and Maloney \cite{freeman_body_2021} & $\checkmark$ & $\checkmark$ White, non-White &  & Interviews \\
Kitson et al. \cite{Kitson2016} & $\checkmark$ & $\checkmark$ Chinese, White &  & Actions \\
Almog et al. \cite{almog2009} &  & $\checkmark$ Arab, non-Arab &  & Actions \\
Martingano et al. \cite{Martingano2022} &  & $\checkmark$ Asian, Black, White &  & Cybersickness \\
Teoh and Cyril \cite{teoh2008exploring} &  & $\checkmark$ Chinese, Malay &  & Presence \\
Scheibler and Rodrigues \cite{scheiblerUser2019} & $\checkmark$ &  &  & SoE \\
Dirin et al. \cite{dirinGender2019} & $\checkmark$ &  &  & Enthusiasm \\
Phosardl et al. \cite{phosaard2010effects} & $\checkmark$ &  &  & Actions \\
Vila et al. \cite{vilaGender2003} & $\checkmark$ &  &  & Actions \\
Zibrek et al. \cite{zibrekEffect2020} & $\checkmark$ &  &  & Actions \\
Sagnier et al. \cite{sagnier2020effects} & $\checkmark$ &  &  & Presence, Actions \\
Schwind et al. \cite{schwind2017} & $\checkmark$ &  &  & Presence \\
Grassini and Laumann \cite{grassiniLaumann2020} & $\checkmark$ &  &  & Cybersickness \\
Munafo et al. \cite{munafoVirtual2017} & $\checkmark$ &  &  & Cybersickness \\
Peck et al. \cite{peck2020gap} & $\checkmark$ &  &  & Cybersickness \\
Shafer et al. \cite{shaferFactors2019} & $\checkmark$ &  &  & Cybersickness \\
Stanney et al. \cite{Stanney2020} & $\checkmark$ &  &  & Cybersickness \\
Xia et al. \cite{xiaColour2022} & $\checkmark$ &  &  & Cognitive performance \\
Peck and Gonzalez-Franco \cite{peck2021} &  &  & Age & Presence \\
Serino et al. \cite{serino2018role} &  &  & Age & Presence \\
Plechata et al. \cite{plechataAgeRelated2019} &  &  & Age & Memory \\
Taillade et al. \cite{tailladeExecutive2013} &  &  & Age & Memory \\ 
Olson \cite{olson2019exploring} &  &  & RES & Actions \\
\hline
\textbf{Ours} & \textbf{$\checkmark$} & \textbf{$\checkmark$ Asian, Black, Hispanic, White} & \textbf{} & \textbf{SoE} \\ \hline
\end{tabular}
\end{table*}

\section{Related Work}
In this section, we provide a brief overview of self-avatars, avatar similarity and dissimilarity effects, and effects of user demographics on VR experiences. 

\subsection{Self-Avatars in Virtual Reality}
The representation of the body through avatars, or self-avatars, holds importance for users \cite{Vasalou2007}, particularly in immersive technologies like VR \cite{Steed2016}. We define self-avatars as avatars wherein "users are embodied by a virtual avatar that is co-located with the user's body" \cite{gonzalez-franco2020b}, and embodiment refers to the phenomenon where users "coexist with a virtual avatar, experiencing it from a first-person perspective" \cite{peck2021}. In VR, where users cannot see their actual bodies, self-avatars become crucial for interactions within the virtual environment. Gonzalez-Franco et al. \cite{gonzalez-franco2020b} emphasize that avatars are central not only in social VR interactions but also in non-social contexts.

Embodiment research explores how users perceive and engage with their self-avatars \cite{spanlang2014}. Spanlang et al. \cite{spanlang2014} argue that virtual embodiment is a valuable tool, particularly in fields where high ecological validity is essential, such as neuroscience and psychology. We are particularly interested in SoE, and it is proposed that SoE is a fundamental aspect of self-consciousness, primarily shaped by the processing of multi-sensory information \cite{roth2020}. This aspect of self-consciousness can serve various research purposes, including facilitating perspective-taking abilities \cite{chen2021, peck2018, peck2020}.

\subsection{Effects of Matching Self-Avatar Appearance}
The effects of matching or not matching avatar appearance can have significant implications for users in virtual immersive environments. Scholars have extensively explored the impact of matching and not matching self-avatar appearance to users, recognizing avatar appearance as a crucial factor in inducing SoE \cite{fribourg_avatar_2020}. Early studies found that the degree of similarity between a user's real and virtual body can impact VR experiences \cite{Jo2017, Jung2018}. Several scholars have looked at how embodying avatars of different races \cite{peck2013, salmanowitz_impact_2018, banakou2016, marini_i_2022, ambron2022, kilteni2013drumming} and genders \cite{peck2018, peck2020} can affect actions and biases. However, most of these studies did not measure SoE, although several studies found that embodying avatars of a different gender did not impact SoE \cite{do2024stepping, peck2018, peck2020, lopez2019}.

Despite the relevance of avatar matching studies, Cheymol et al. \cite{cheymol_beyond_2023} emphasize the scarcity of studies particularly analyzing the \textit{degree} of similarity and dissimilarity between avatars and users, arguing that current research lacks a comprehensive understanding and methodologies in this area. Additionally, Do et al. \cite{do2024stepping} provided a comprehensive overview of studies that investigated matching user characteristics to their avatar's, and found that a substantial number of studies require users to embody White avatars as a condition, regardless of the user's ethnicity, which can negatively affect SoE. Of particular relevance to our work is recent research highlighting the importance of matching an avatar's ethnicity to the user's in VR, which can improve SoE \cite{do2024stepping} and self-identity \cite{freeman_body_2021}. However, these prior works did not formally evaluate the effects of ethnicity-matched avatars on different ethnicity groups. We extend the experimental design of Do et al. \cite{do2024stepping} while explicitly controlling for participant ethnicity and gender and recruiting a sufficient sample size.

\subsection{Effects of User Demographics on VR Experiences}
User demographics can have significant effects on VR experiences in general, although this area of study is relatively scarce. However, the importance of researching this area is increasing as VR becomes more popular worldwide, and scholars have argued for further exploration in this field \cite{peck2021divrsify}. Peck et al. \cite{peck2021divrsify} in particular have provided an excellent review of how participant demographics may differ in VR experiences, spanning factors such as gender, race or ethnicity, and age \cite{serino2018role, peck2021}.  

Much work in this area has focused on gender, which appears to be a significant factor influencing VR experiences, as indicated by a substantial body of research. In a study on the impact of gender on the SoE, Scheibler and Rodrigues \cite{scheiblerUser2019} found that women and men differed in their SoE based on the point of view, although avatar race was fixed. Other studies have reported that men tend to exhibit higher levels of presence in VR compared to women \cite{felnhofer2012virtual, teoh2008exploring}. Similarly, Schwind et al. \cite{schwind2017} noted that women expressed a dislike for male hands and demonstrated lower levels of presence, while men did not exhibit differences based on hand genders. Gender differences extend to distinct actions, as observed in various studies \cite{phosaard2010effects, zibrekEffect2020, vilaGender2003, sagnier2020effects}. Peck et al. \cite{peck2020gap} emphasized the underrepresentation of female participants in VR studies and highlighted a correlation between the proportion of female participants and simulator sickness. They advocated for increased representation of female participants in VR research to address potential bias in findings. Considering these insights, we made efforts to recruit an approximately equal number of male and female participants, incorporating gender as a between-subjects factor in our study design and analyses.

Several studies have explored the influence of user ethnicity, race, or culture on VR experiences. Teoh and Cyril \cite{teoh2008exploring} observed that Malay participants reported higher spatial presence compared to Chinese participants. Further investigations into how ethnicity influences actions in VR have been conducted. Almog et al. \cite{almog2009} found that Arab women were less likely to look out of a virtual window than non-Arab women, suggesting that ethnicity or culture can affect the sense of presence. Kitson et al. \cite{Kitson2016} reported that White participants were more likely to turn their heads in response to simulated heading changes compared to Chinese participants. Similarly, Olson \cite{olson2019exploring} found that participants' prior experiences with racial and ethnic socialization influenced their actions in a VR simulation about racial discrimination. However, research in this area remains scarce \cite{peck2021divrsify}, and there have been limited investigations into how user ethnicity affects SoE. We expand upon these studies by exploring how participant ethnicity influences avatar matching effects to gain insights into how users may perceive themselves in social immersive environments. 

\subsection{Personalized Avatars}
Recent studies, such as those by Kim et al. \cite{kimBe2023}, have demonstrated that personalized avatars significantly enhance the sense of embodiment compared to random avatars of the same race and gender. Foundational work by Waltemate et al. \cite{waltemate2018} and Salagean et al. \cite{salagean2023} showed that avatars personalized through photogrammetry lead to a stronger sense of body ownership than generic gender-matched avatars.

However, some studies indicate that avatar realism interacts with personalization. Jo et al. \cite{jo_impact_2017} found participants felt more embodied in a generic avatar with similar clothing compared to a realistic point-cloud representation. Dollinger et al. \cite{dollinger2023} observed that personalized photorealistic avatars, while increasing the sense of embodiment over customized and generic avatars, also resulted in higher eeriness and reduced body awareness. Despite these findings, many personalization methods remain currently impractical for consumer applications and some research environments.

\section{Methods}

\begin{figure}[h!]
    \centering
    \includegraphics[width=2in]{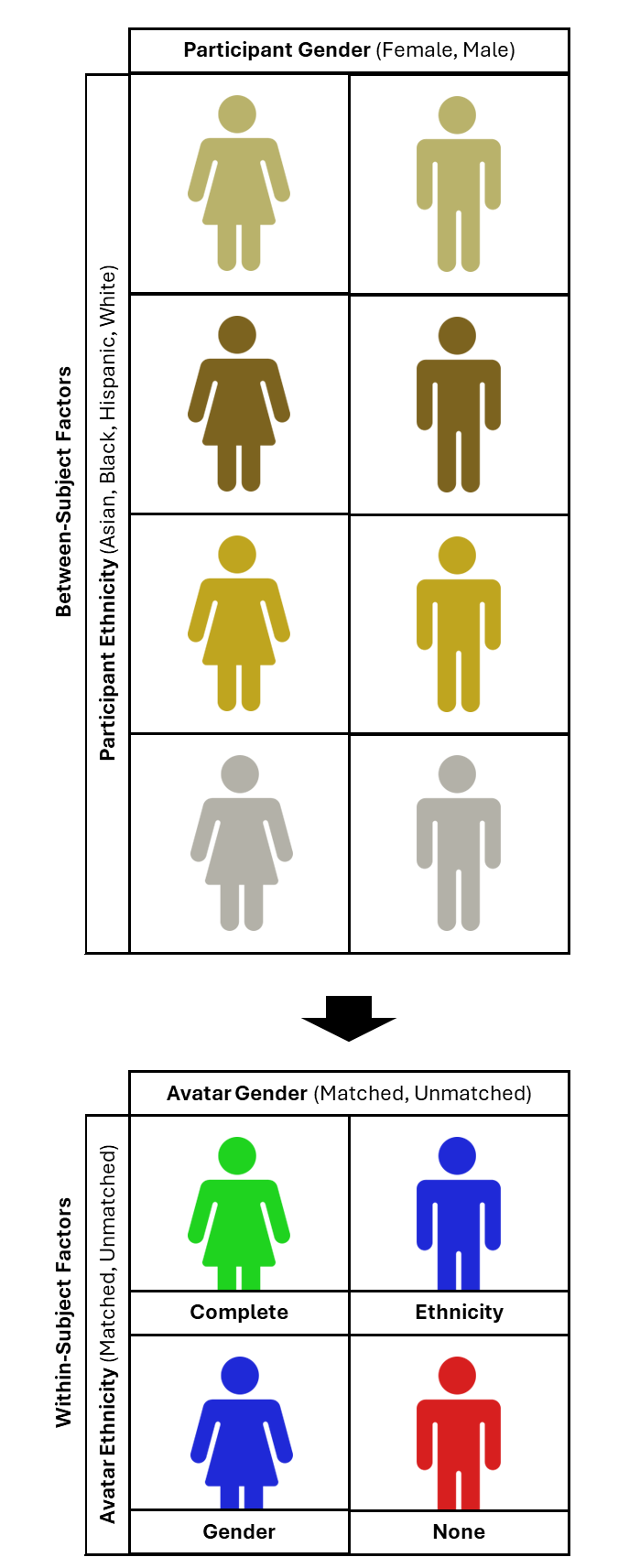}
    \caption{Visualization of our study design factors.}
    \label{fig:studydesign}
\end{figure}

We conducted a within-subjects 2 (matched or unmatched avatar ethnicity) x 2 (matched or unmatched avatar gender), and between-subjects (participant ethnicity: Asian, Black, Hispanic, or White) x 2 (participant gender: male or female) mixed model experiment to assess the impact of matching or not matching a user's self-avatar with regards to their ethnicity and/or gender. See Figure \ref{fig:studydesign} for a depiction of our study design and factors. 

\begin{figure}[h!]
    \centering
    \includegraphics[width=2.5in]{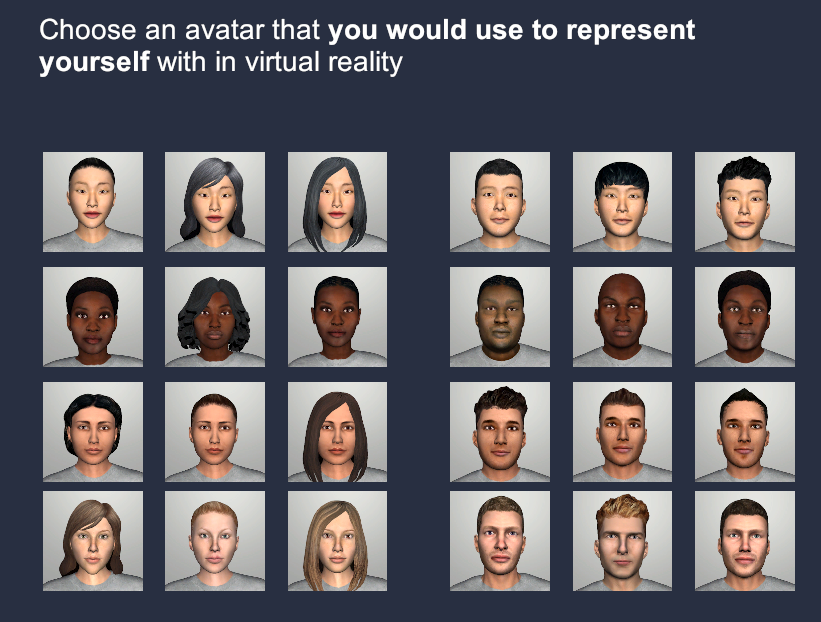}
    \caption{Images of avatars participants were allowed to choose from. For each condition, they were allowed to choose from three avatars. For example, a Black female participant would be able to choose from the three Black female avatars for the \textit{Complete} condition.}
    \label{fig:selection}
\end{figure}

Participants embodied four avatars in a Latin square counterbalanced order, each characterized by their self-reported ethnicity and gender: \textit{Complete} (same ethnicity, same gender), \textit{Ethnicity} (same ethnicity, different gender), \textit{Gender} (different ethnicity, same gender), and \textit{None} (different ethnicity, different gender). To mitigate ordering effects by ethnicity, the counterbalanced order was specific to each participant's ethnicity (i.e., Latin square ordering was respective to ethnicity). For each condition, participants were provided with the option to choose from three avatars (see Figure \ref{fig:selection} for an example).

\begin{figure}[h!]
    \centering
    \includegraphics[width=3.2in]{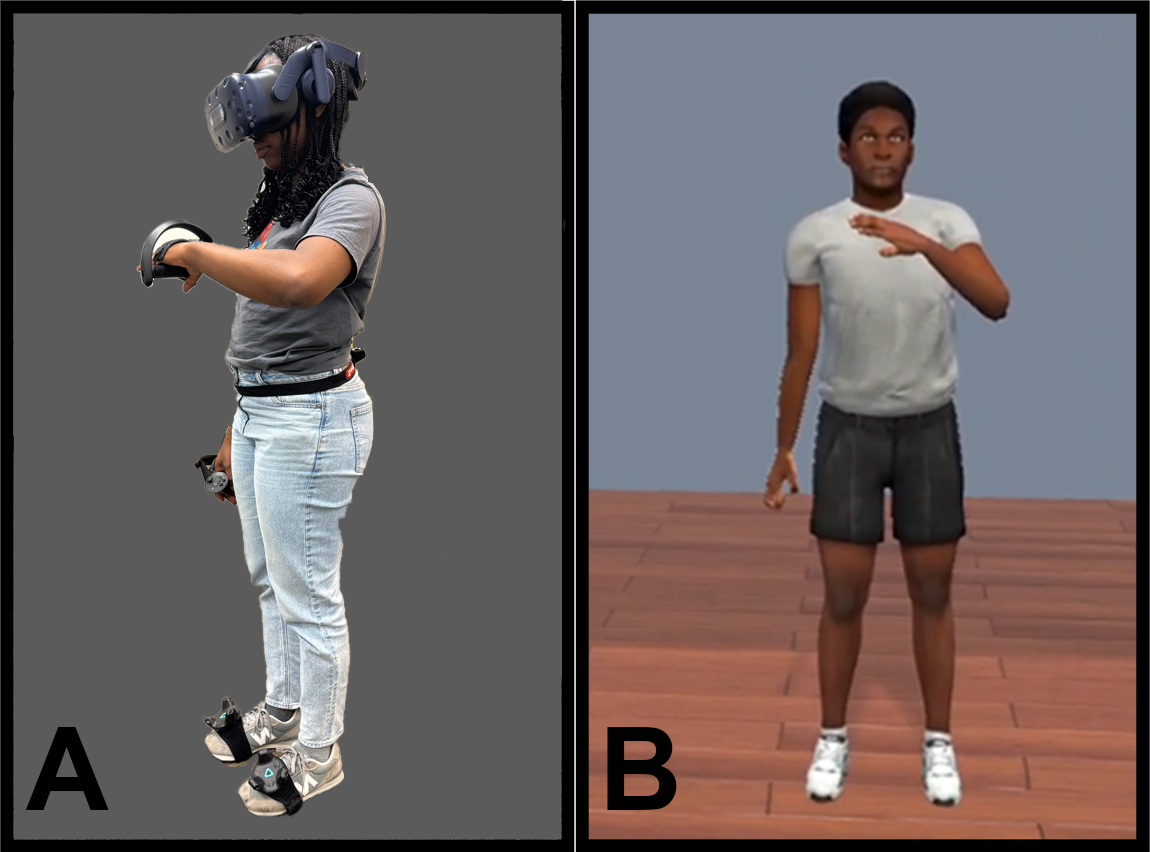}
    \caption{A) The apparatus and tracker placement on a participant. Two trackers are positioned on the feet, and another tracker is placed on the back. B) Image of the participant's reflection in the virtual mirror. In this image, the participant is embodying the \textit{Ethnicity} condition, where the avatar matches her ethnicity but not her gender.}
    \label{fig:apparatus}
\end{figure}

We utilized avatars sourced from the \textit{Virtual Avatar Library for Inclusion and Diversity} (\textit{VALID}) \cite{do2023valid}, which offers avatars validated to be perceived as specific ethnicities and genders. For the ``different gender" avatars, participants embodied an avatar of the opposite gender (female or male). For the ``different ethnicity" avatars, we selected avatars that were least likely to be perceived as the respective ethnicity, based on agreement rates from VALID like Do et al. \cite{do2024stepping}. Consequently, for the different ethnicity conditions, White and Asian participants embodied Black avatars, while Black participants embodied White avatars. Hispanic participants embodied Asian avatars. All avatars wore a plain gray shirt, black shorts, and white shoes, representing the ``Default" outfits from the VALID library (see Figure \ref{fig:apparatus}). 

\subsection{Research Hypotheses}
We had the following hypotheses for our research questions:

    \textbf{H1a:} Participants will report higher SoE when avatar ethnicity is matched in comparison to unmatched ethnicity \cite{do2024stepping}.

    \textbf{H1b:} Participants will have no significant differences when avatar gender is matched or unmatched \cite{do2024stepping}. 
    
    \textbf{H2a:} Non-White participants will experience a lower SoE when embodying a different ethnicity compared to White participants. This hypothesis is based on previous interview findings suggesting that skin tone holds particular significance for non-White participants \cite{freeman_body_2021}. 

    \textbf{H2b:} No differences are anticipated between men and women concerning avatar matching effects. However, we anticipate that women will have lower SoE than men \cite{scheiblerUser2019}.

\subsection{Dependent Variables}

After completing each condition, we administered standardized embodiment questionnaires, as detailed below. Additionally, participants were asked to complete an exit survey after finishing all conditions to collect qualitative information.

\subsubsection{Standardized Embodiment Questionnaire (SEQ)}
We used the 2021 Standardized Embodiment Questionnaire (SEQ) developed by Peck and Gonzalez-Franco \cite{peck2021} to analyze SoE. The SEQ consists of sixteen 7-point Likert-scale questions. The SEQ produces a total Embodiment score, along with four subscales: Appearance, Response, Ownership, and Multi-Sensory.

\subsection{Apparatus}
Participants utilized an HTC Vive Pro accompanied by three additional VIVE trackers placed on their feet and lower back. This tracker configuration closely aligns with one of the most widely adopted consumer setups for full-body tracking in applications like VRChat or Dance Rush \cite{krell2023corporeal}. For hand and finger tracking, participants used Valve Index controllers. All movements were replicated from these devices to the participants' avatars through an inverse kinematics API (FinalIK\footnote{http://root-motion.com/}). The application was developed using Unity and SteamVR platforms. Participants found themselves in a virtual bedroom, positioned approximately 1.5 meters from a virtual mirror.

\subsection{Procedure}
The study consisted of one in-person session that lasted approximately 45 minutes and received approval from an Institutional Review Board (IRB). Participants first completed an online screener that captured their demographics, including their ethnicity, gender, education, and VR experience. 

Upon arrival to the lab, participants were assigned to one of four Latin squares ordering cohorts for counterbalancing the four within-subject conditions. Subsequently, they were instructed to choose four characters, each corresponding to one of the experimental conditions, using an avatar selection interface displayed on a computer. For example, for the "Ethnicity" condition,  participants were presented only with avatars of their matched ethnicity, but not their gender.  Participants were able to choose from three different characters for each condition. Participants were only given the instructions, ``Choose an avatar that you would use to represent yourself in virtual reality" for each selection.

Following the character selection, participants donned the VR headset and controllers. Participants were then assisted in putting on the VIVE trackers. Upon wearing all equipment, the application initiated the avatar calibration process, which adjusted the avatar's features based on the user's height. The avatar remained concealed until the calibration was finalized. Subsequently, participants engaged in a standard embodiment procedure in front of a virtual mirror. This embodiment procedure was developed by Roth and Latoschik \cite{roth2020} and was adapted by Do et al. \cite{do2024stepping} to include additional instructions pertaining to feet and object interactions. Upon concluding the task within the virtual environment, participants exited the VR experience and proceeded to complete the Sense of Embodiment Questionnaire (SEQ) on a computer.

After completing all four conditions, participants participated in a brief exit interview to allow them to express free response thoughts. In this exit interview, participants were given the following two questions, with no follow up questions: 1) ``Which avatar(s) did you feel most embodied by? Why?'' 2) ``Which avatar(s) did you feel least embodied by? Why?''

\subsection{Collected Data}
Alongside our dependent variables, we also collected a set of motion data for each avatar (e.g., \textit{Complete}, \textit{Ethnicity}, \textit{Gender}, and \textit{None}) for each participant. This set of anonymized data includes the local tracking position $(x, y, z)$ and rotation (Euler angles, quaternions, and 6D representations) of all VR tracked devices (headset, controllers, foot trackers, and back tracker) every frame at 90Hz. This motion data was collected to potentially explore how various demographics move in VR environments for potential future research. Participants were duly informed about the sharing of their data, encompassing demographics and motion data, and they provided explicit consent through a signed consent form. This release of data was formally approved by an IRB.

\begin{table}[h!] \centering
\caption{Demographics of all participants who completed the study, organized by participant ethnicity. The table includes gender (Female or Male), age, ownership of a consumer VR system (Yes or No).}
\begin{tabular}{@{}llll@{}}
\toprule
\textbf{Ethnicity} & \textbf{Gender} & \textbf{Age M(SD)} & \textbf{VR Ownership} \\ \midrule
Asian              & 10M, 10F        & 21.90 (3.92)       & 4Y, 16N               \\
Black              & 10M, 8F         & 21.28 (2.56)       & 4Y, 14N               \\
Hispanic           & 10M, 10F        & 20.05 (1.81)       & 5Y, 15N               \\
White              & 11M, 9F         & 21.35 (3.67)       & 4Y, 16N               \\ \bottomrule
\end{tabular}
\label{table:particiapnts}
\end{table}

\begin{table*}[ht!] \centering
\setlength\extrarowheight{3pt} 
\caption{Mean and standard deviation measurements of all participants, organized by participant ethnicity and avatar condition. Participant gender was not disaggregated since results showed that participant gender was not a significant factor for any measurement. }
\resizebox{\textwidth}{!}{%
\begin{tabular}{lccccc|c|lccccc}
\hline
 & \textbf{SoE} & \textbf{Appear.} & \textbf{Response} & \textbf{Ownership} & \textbf{Multi-S.} &  &  & \textbf{SoE} & \textbf{Appear.} & \textbf{Response} & \textbf{Ownership} & \textbf{Multi.S} \\ \hline
\multicolumn{6}{c|}{\textit{\textbf{Asian participants (n=20)}}} &  & \multicolumn{6}{c}{\textit{\textbf{Black participants (n=18)}}} \\ \hline
\textbf{Complete} & 3.06 (1.01) & 2.75 (1.15) & 2.54 (1.04) & 3.47 (1.22) & 3.46 (1.00) &  & \textbf{Complete} & 3.10 (0.96) & 2.94 (1.18) & 2.52 (1.17) & 3.67 (0.75) & 3.26 (1.01) \\
\textbf{Ethnicity} & 2.92 (1.00) & 2.71 (1.29) & 2.29 (1.11) & 3.24 (1.04) & 3.45 (1.00) &  & \textbf{Ethnicity} & 2.85 (0.81) & 2.71 (1.12) & 2.17 (1.04) & 3.29 (0.57) & 3.22 (0.79) \\
\textbf{Gender} & 2.81 (1.04) & 2.62 (1.14) & 2.32 (1.10) & 2.89 (1.05) & 3.41 (1.21) &  & \textbf{Gender} & 2.61 (0.96) & 2.64 (1.23) & 2.08 (1.00) & 2.81 (0.85) & 2.90 (1.06) \\
\textbf{None} & 2.69 (1.18) & 2.52 (1.30) & 2.28 (1.30) & 2.81 (1.17) & 3.14 (1.21) &  & \textbf{None} & 2.54 (0.96) & 2.57 (1.15) & 2.07 (1.12) & 2.63 (0.91) & 2.90 (1.03) \\ \hline
\multicolumn{6}{c|}{\textit{\textbf{Hispanic participants (n=20)}}} &  & \multicolumn{6}{c}{\textit{\textbf{White participants (n=20)}}} \\ \hline
\textbf{Complete} & 3.69 (1.35) & 3.53 (1.64) & 3.26 (1.44) & 3.94 (1.16) & 4.05 (1.41) &  & \textbf{Complete} & 3.45 (0.76) & 3.10 (0.97) & 2.77 (0.80) & 3.90 (0.84) & 4.04 (1.01) \\
\textbf{Ethnicity} & 3.61 (1.09) & 3.62 (1.40) & 3.15 (1.27) & 3.65 (0.93) & 4.01 (1.10) &  & \textbf{Ethnicity} & 3.15 (0.78) & 2.95 (0.91) & 2.44 (0.80) & 3.40 (0.85) & 3.81 (1.03) \\
\textbf{Gender} & 3.68 (1.15) & 3.69 (1.43) & 3.12 (1.29) & 3.91 (1.04) & 4.11 (1.16) &  & \textbf{Gender} & 3.15 (0.74) & 2.97 (0.92) & 2.44 (0.89) & 3.29 (0.67) & 3.91 (1.04) \\
\textbf{None} & 3.71 (1.37) & 3.69 (1.71) & 3.32 (1.52) & 3.64 (1.28) & 4.19 (1.33) &  & \textbf{None} & 3.28 (0.64) & 3.13 (0.71) & 2.49 (0.70) & 3.39 (0.73) & 4.09 (1.03) \\ \hline
\end{tabular}}
\label{table:byethnicity}
\end{table*}

\subsection{Participants}
A total of 78 participants were recruited from a university and surrounding areas via listservs and flyers. All participants reported normal or corrected-to-normal vision. Participant demographics, including gender, ethnicity, and age, can be found in Table \ref{table:particiapnts}.

\subsection{Data Analysis Approach}
Following the guidelines provided by \cite{peck2021}, we computed the averages of all subscale questions and the total embodiment score. Subsequently, we conducted a Shapiro-Wilk test to determine the normality of the score distributions. Given that this test revealed non-normal distributions for several measures, we used the Aligned Rank Transform (ART) method \cite{Wobbrock2011} for 2 $\times$ 2 $\times$ 4 $\times$ 2 factorial analysis of variance (ANOVA). Using this procedure, we tested for interaction effects and main effects of matched avatar ethnicity (matched or unmatched), matched avatar gender (matched or unmatched), participant ethnicity (Asian, Black, Hispanic, and White), and participant gender (male or female). We also performed post-hoc Aligned Rank Transform Contrast (ART-C) \cite{Elkin2021} tests for pairwise comparisons. We used an alpha level of 0.05, and for post-hoc tests, we used a Holm-Bonferroni adjustment.

\section{Results}
In this section, we first present the findings related to the total SoE score. Subsequently, we elaborate on the results concerning the subscales. Table \ref{table:overall} shows the data of all participants, organized by avatar condition. Table \ref{table:byethnicity} shows measurements of all participants, organized by participant ethnicity and avatar condition. Participant gender was not disaggregated since results showed that participant gender was not a significant factor for any measurement. The results of all statistical tests can be found in Appendix A.


\begin{table}[h!]
\caption{Mean and standard deviation measurements of all participants, organized by avatar condition}
\resizebox{\columnwidth}{!}{%
\begin{tabular}{@{}lccccc@{}}
\toprule
 & \textbf{SoE} & \textbf{Appear.} & \textbf{Response} & \textbf{Ownership} & \textbf{Multi-S} \\ \midrule
\textbf{Complete} & 3.33 (1.06) & 3.08 (1.27) & 2.78 (1.16) & 3.74 (1.02) & 3.72 (1.16) \\
\textbf{Ethnicity} & 3.14 (0.96) & 3.00 (1.23) & 2.52 (1.12) & 3.40 (0.87) & 3.63 (1.02) \\
\textbf{Gender} & 3.07 (1.05) & 2.96 (1.24) & 2.50 (1.13) & 3.23 (1.00) & 3.60 (1.19) \\
\textbf{None} & 3.07 (1.15) & 2.99 (1.34) & 2.55 (1.27) & 3.13 (1.11) & 3.60 (1.27) \\ \bottomrule
\end{tabular}
\label{table:overall}}
\end{table}

\subsection{Total SoE}

\begin{figure}[h!]
    \centering
    \includegraphics[width=\columnwidth]{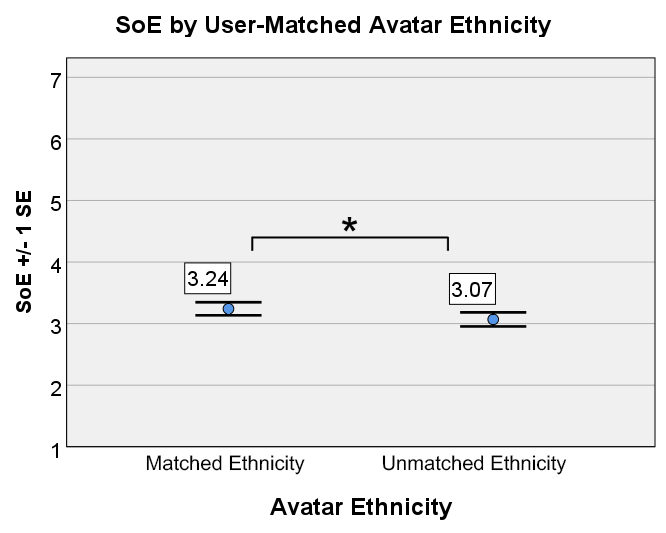}
    \caption{Plot of total SoE scores, separated by user-matched avatar ethnicity. An asterisk (*) denotes a significant difference between conditions. Error bars represent +/- 1 standard error.}
    \label{fig:graph_soeethnicity}
\end{figure}

\begin{figure}[h!]
    \centering
    \includegraphics[width=\columnwidth]{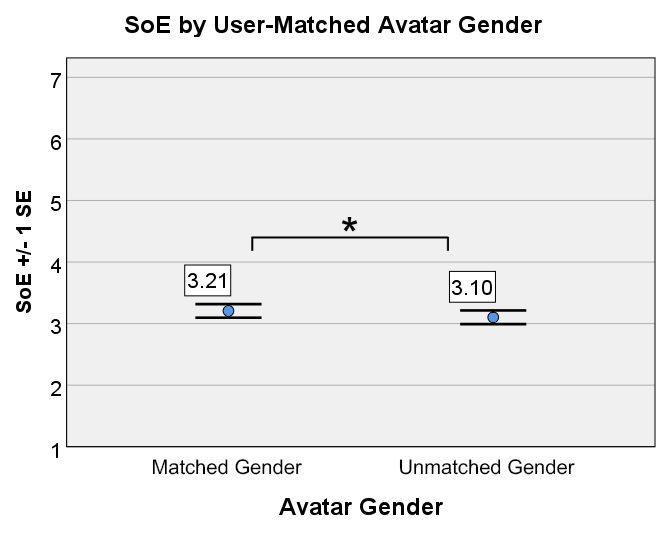}
    \caption{Plot of total SoE scores, separated by user-matched avatar gender. An asterisk (*) denotes a significant difference between conditions. Error bars represent +/- 1 standard error.}
    \label{fig:graph_soegender}
\end{figure}

No significant interaction effects were found between any of the factors. However, significant main effects were observed for matched ethnicity (see Figure \ref{fig:graph_soeethnicity}) and matched gender (see Figure \ref{fig:graph_soegender}), respectively. Participants reported a higher SoE when embodying an avatar of the same ethnicity, $F(1, 70) = 10.21, p < 0.01, \eta^2=0.13$, and when embodying an avatar of the same gender $F(1, 70) = 4.78, p = 0.03, \eta^2=0.06$. Thus, H1a was supported, while H1b was not supported, considering that we found significant differences of matched gender.

\begin{figure}[h!]
    \centering
    \includegraphics[width=\columnwidth]{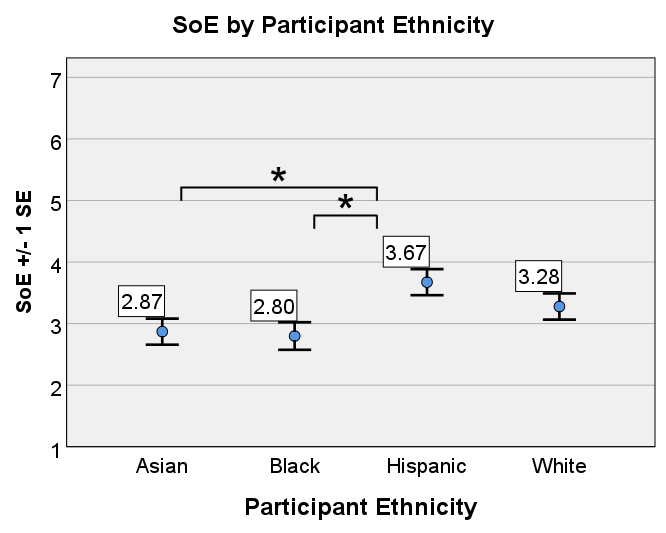}
    \caption{Plot of total SoE scores, separated by participant ethnicity. An asterisk (*) denotes a significant difference between groups. Error bars represent +/- 1 standard error.}
    \label{fig:graph_soepethnicity}
\end{figure}

Surprisingly, a main effect of participant ethnicity was identified, $F(3, 70)=4.16, p<0.01, \eta^2=0.15$ (see Figure \ref{fig:graph_soepethnicity}). Post-hoc tests indicated that Asian participants had significantly lower SoE scores than Hispanic participants, $t(70) = -2.71, p = 0.04$. Similarly, Black participants had significantly lower SoE scores than Hispanic participants, $t(70) = -2.94, p = 0.02$. No other main effects were observed. Since there were no differences between men and women, H2b was not supported.

\subsection{Subscales}

\subsubsection{Appearance}
We did not find any significant main effects or interaction effects for the \textit{Appearance} subscale.

\subsubsection{Response}
\begin{figure}[h!]
    \centering
    \includegraphics[width=\columnwidth]{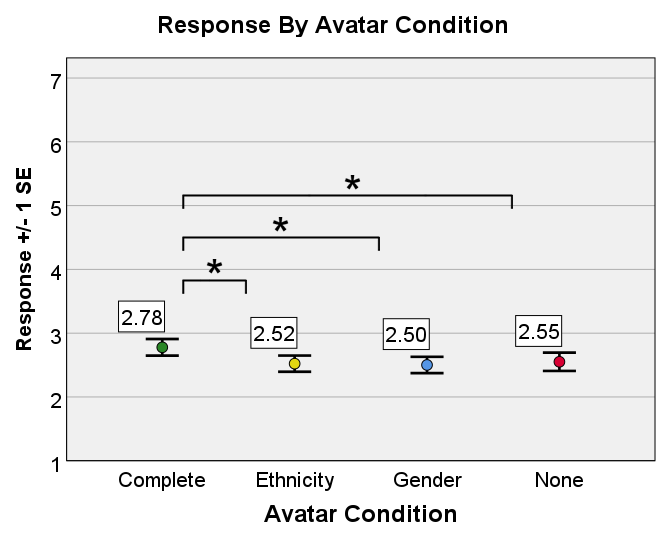}
    \caption{Plot of \textit{Response} subscale scores, separated by user-matched avatar conditions (Complete, Ethnicity, Gender, and None). An asterisk (*) denotes a significant difference between conditions. Error bars represent +/- 1 standard error.}
    \label{fig:response}
\end{figure}

A significant interaction effect was observed between matched avatar ethnicity and matched avatar gender for the \textit{Response} subscale, $F(1, 70) = 6.04, p = 0.02, \eta^2=0.07$ (see Figure \ref{fig:response}). Post-hoc tests indicated that the \textit{Complete} condition had higher \textit{Response} scores than all other conditions: the \textit{Ethnicity} condition, $t(70)=3.01, p = 0.01$, the \textit{Gender} condition, $t(70)=3.24, p < 0.01$, and the \textit{None} condition, $t(70)=3.11, p=0.01$.  No other interaction effects or main effects were found for this subscale.

\begin{figure}[h!]
    \centering
    \includegraphics[width=\columnwidth]{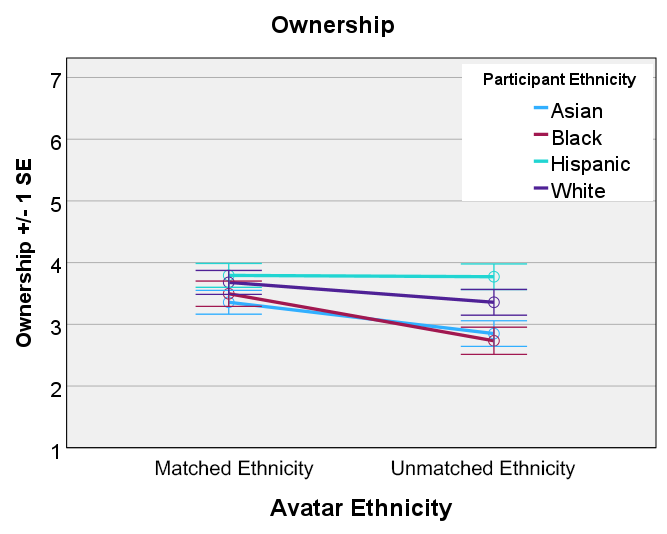}
    \caption{Interaction plot of \textit{Ownership} subscale scores, separated by user-matched avatar ethnicity conditions and participant ethnicity. Error bars represent +/- 1 standard error.}
    \label{fig:graph_ownership}
\end{figure}

\begin{figure}[h!]
    \centering
    \includegraphics[width=\columnwidth]{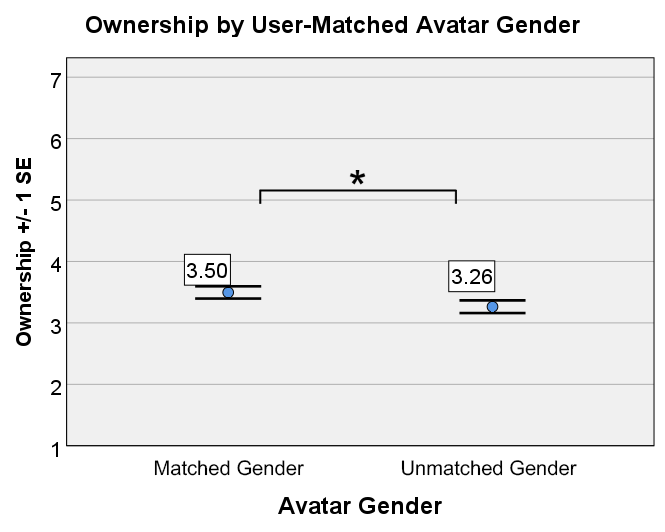}
    \caption{Plot of \textit{Ownership} subscale scores, separated by user-matched avatar gender. An asterisk (*) denotes a significant difference between conditions. Error bars represent +/- 1 standard error.}
    \label{fig:graph_ownership2}
\end{figure}

\subsubsection{Ownership}

A single significant interaction effect was observed between matched avatar ethnicity and participant ethnicity for the \textit{Ownership} subscale, $F(3, 70)=6.25, p < 0.01, \eta^2=0.21$. Figure \ref{fig:graph_ownership} illustrates a plot of \textit{Ownership} scores, separated by participant ethnicity. 

This interaction effect prompted us to disaggregate data by participant ethnicity to shed further insights. Upon doing so, we found that different groups exhibited substantial differences in how their \textit{Ownership} was affected by matching avatar characteristics.As depicted in Figure \ref{fig:graph_ownership}, Hispanic participants exhibited no significant differences in \textit{Ownership} between avatars that matched their ethnicity and avatars that did not match their ethnicity, $F(1, 18)=0.16, p = 0.68, \eta^2<0.01$. White participant had significantly higher \textit{Ownership} scores when avatars matched their ethnicity, $F(1, 18)=5.50, p = 0.02, \eta^2=0.23$. However, Asian participants, $F(1, 18)=17.37, p<0.01, \eta^2=0.49$, and Black participants, $F(1, 16)=42.44, p < 0.01, \eta^2=0.73$, had significantly higher \textit{Ownership} scores with greater effect sizes than White participants. These results partially support H2a, as Asian and Black participants were more affected by unmatched ethnicity avatars. 

No additional interaction effects were identified. However, a significant main effect of matched avatar gender was observed, indicating that participants experienced higher \textit{Ownership} when embodying an avatar of the same gender, $F(1, 70)=17.51, p<0.01, \eta^2=0.08$. No other main effects were detected.

\subsubsection{Multi-Sensory}
Interestingly, we found a main effect of participant ethnicity for the \textit{Multi-Sensory} subscale, $F(3,70)=4.90, p < 0.01, \eta^2=0.17$. Subsequent post-hoc tests revealed that Black participants exhibited significantly lower levels of \textit{Multi-Sensory} feelings than both White participants, $t(70)=-2.93, p=0.02$, and Hispanic participants, $t(70)=-3.28, p<0.01$. There were no other main or interaction effects. 

\begin{figure}[h!]
    \centering
    \includegraphics[width=\columnwidth]{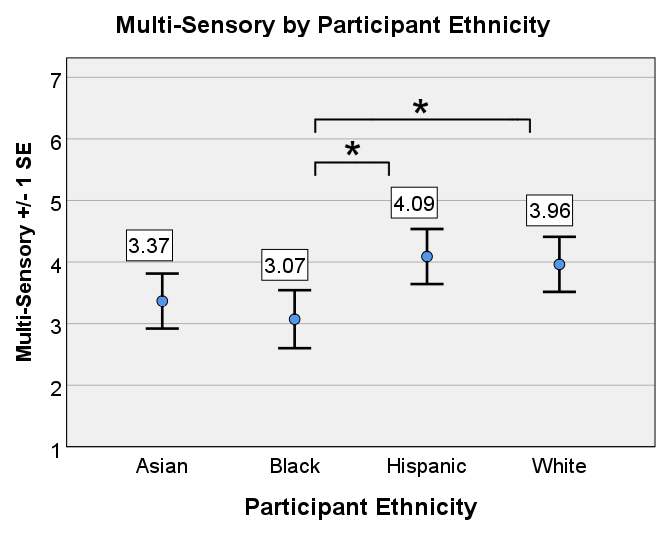}
    \caption{Plot of Multi-Sensory subscale scores, separated by participant ethnicity. An asterisk (*) denotes a significant difference between groups. Error bars represent +/- 1 standard error.}
    \label{fig:graph_multisensory}
\end{figure}

\section{Discussion}
In this section, we discuss the impact of user-matched avatars on the sense of embodiment, followed by an examination of how user ethnicity interacts with these matching effects. We delve into the ways our findings illuminate the influence of user ethnicity on various aspects of embodiment. Throughout our discussion, we integrate participant quotes from post-study interviews to offer deeper insights that complement our quantitative results. Finally, we conclude this section by addressing the limitations of our study and proposing potential avenues for future research. This paper makes the following contributions: 
\begin{enumerate}
    \item We provide novel insights on how user racial-ethnic identity may affect SoE as well as multi-sensory perceptions. Specifically, we found that Asian and Black participants generally had lower SoE. 
    \item Our findings highlight a significant interaction effect of user ethnicity for ownership, indicating that Asian and Black participants are more impacted by avatars of different ethnicities than White or Hispanic participants.
    \item We provide results demonstrating the importance of matching a user's ethnicity and gender, particularly depending upon the user's racial-ethnic identity.
    \item In an effort to increase diversity in VR datasets, we provide our dataset, which includes the tracked movements and demographics of all 78 participants in the study. This dataset can be found at: \url{https://doi.org/10.17605/OSF.IO/YDW48}. 
\end{enumerate}

\subsection{User Ethnicity Affects SoE and Multi-Sensory}
Our study uncovered a noteworthy main effect of participant ethnicity on total SoE and the Multi-Sensory subscale, which we did not initially anticipate in our original hypotheses. Specifically, Asian and Black participants reported lower SoE than Hispanic participants, and Black participants exhibited lower scores on the Multi-Sensory subscale compared to both Hispanic and White participants. This exploration into distinctions among various racial-ethnic demographics in VR use contributes to the limited body of empirical research in this domain, and we urge the community to further explore such effects.

Crucially, these results did not exhibit interaction effects with matched/unmatched avatar characteristics. The influence of participant ethnicity extended across all conditions, indicating that Asian and Black participants had lower SoE regardless of avatar congruency.

While our study did not delve deeply into the reasons behind these results, it raises questions about the design focus for these populations in VR experiences. Previous research suggests that VR may be less enjoyable for women due to headsets being initially designed and tested on men \cite{Stanney2020}. Our findings suggest that future work should investigate reasons why Asian and Black participants may experience lower SoE. Possible explanations may include issues with inverse kinematics (IK) not properly aligning with their movements. Participants noted a lack of embodiment, with one expressing that it \textit{"just did not feel right"} (Black, M). A participant even noted:
\begin{quote}
    \textit{"I felt like I embodied none of them because they all felt the same level of unrealistic due to their movements."} (Asian, M)
\end{quote}

We hypothesize that a user's ethnic-racial identity may impact their movement in VR, and generic consumer IK solutions may not cater well to specific populations. Scholars have argued that everyone moves uniquely in VR  \cite{rack2023, miller2020personal, moore2021, nair2023}. We encourage future researchers and developers to investigate how racial-ethnic identity shapes VR experiences, pinpointing reasons for disparities and striving to enhance these experiences.

\subsection{User Ethnicity Interacts with Avatar Matching Effects}
We found that a user's ethnicity played a significant role in influencing avatar matching effects, specifically for Ownership. In our study, we found that Asian, Black, and White participants all reported higher Ownership when embodying avatars of the same ethnicity as their own. 

Previous interviews suggested that non-White users may place greater importance on their self-avatars' skin tone and ethnic features in VR \cite{freeman_body_2021}. This emphasis on ethnic features was evident among our Asian and Black participants. Specifically, for the effect of matched avatar ethnicity on Ownership, Black participants had an effect size of $\eta^2=0.73$, and Asian participants had an effect size of $\eta^2=0.49$, which can be interpreted as very large effects. In contrast, White participants showed a more moderate effect size of $\eta^2=0.23$. Thus, our results indicate that Asian and Black participants may be more influenced by matching avatar ethnicity compared to White participants. 

We hypothesize that these results may be attributed to racial-ethnic identity. Previous research has indicated that Black, Asian, and Latinx participants tend to have stronger ethnic-racial identities than White participants \cite{searsOrigins2003, vargasDocumenting2016}. For instance, an early study by Sears et al. \cite{searsOrigins2003} found that non-White college students exhibited significantly stronger ethnic identity than their White counterparts, leading to higher racial awareness, identity salience, and feelings of closeness with their racial group. This research also suggested that individuals from these demographics tend to think about their race and ethnicity more frequently than White students, highlighting the salience of racial-ethnic identity in their lives.

In order to gain some preliminary insight on why participants may have felt this way, we analyzed responses from our post-study interviews. In these interviews, participants of Asian and Black descent expressed shared sentiments regarding the challenges of embodying avatars of a different ethnicity. Many participants highlighted the difficulty of reconciling their mental self-image with an avatar of a distinct ethnicity, potentially explaining the pronounced effects on Ownership. For instance, participants mentioned:

\begin{quote}
    \textit{"Because in the back of my mind, I am fully aware that I do not look like [a different ethnicity] in reality, so it is quite hard to lie to yourself that you are someone totally opposite of what you actually are."} (Asian, F) 
    \\ \\
    \textit{"I felt that my brain could not associate that [different ethnicity] avatar as an extension of myself."} (Asian, M)
\end{quote}

One participant even conveyed a sense of complete detachment, stating:

\begin{quote}
    \textit{"This avatar looked the least like me, so when completing any of the actions, it was more like I was moving through a simulation versus simulating my own movements."} (Black, M)   
\end{quote}

These insights highlight the impact of incongruent avatar ethnicity, providing a nuanced understanding of the challenges faced by Asian and Black participants when embodying avatars misaligned with their ethnic identity.

However, our study unexpectedly revealed that Hispanic participants did not exhibit significant differences in Ownership when embodying avatars of the same race compared to avatars of another ethnicity. One possible explanation for this finding may lie in the dynamic nature of ethnic-racial identities, particularly in the United States. Recent work suggests that ethnic-racial identity may be evolving, particularly for Hispanic participants. Martinez-Fuentes et al. \cite{martinez-fuentesExamination2020} found that exploration and resolution, key components of ethnic-racial identity development theory \cite{eriksonIdentity1968}, showed no positive association with Hispanic adolescents, unlike their Black and White counterparts. These recent trends extend to other psychological effects, such as the association between ethnic-racial identity and lower somatic symptoms in Asian students, which did not hold for Hispanic students \cite{rogers-sirinCultural2012}. These complex dynamics in ethnic-racial identity development may contribute to the observed variations in Ownership scores among Hispanic participants.

Notably, when avatars align with participants' ethnicity, Ownership levels remained consistent across all ethnic groups, highlighting the importance of this alignment. However, deviations emerge when avatars do not match participants' ethnicity, revealing notable discrepancies (see Figure \ref{fig:graph_ownership}). Specifically, Black and Asian participants exhibited significantly lower Ownership scores compared to White and Hispanic participants in such instances. This raises a crucial concern about the potential adverse effects of mismatching avatar ethnicity in studies. These findings suggest that such mismatches may disproportionately impact the SoE of Black and Asian participants, introducing the risk of inequitable study outcomes. Therefore, ensuring proper avatar ethnicity matching is imperative for fostering inclusivity and minimizing bias in VR research.

\subsection{Both Matched Avatar Ethnicity and Gender Affect SoE}
In comparison to previous work by Do et al. \cite{do2024stepping}, our study, employing the same procedure and Sense of Embodiment questionnaire, yielded different results. While they found that matched avatar ethnicity influenced the total SoE score and matched avatar gender did not, our findings indicate that both matched avatar gender and matched avatar ethnicity emerged as significant main effects. This discrepancy could be attributed to our much larger sample size (78 participants compared to their 32). 

However, we observed a notable interaction effect for the Response subscale, revealing that an ethnicity-matched, gender-matched avatar elicited a higher Response score than the combined influence of these factors individually. The Response subscale predominantly addresses participants' perceptions of how their real body transforms into their virtual body, incorporating questions like ``It felt as if my (real) body were turning into an `avatar' body'' \cite{peck2021}. While both avatar gender and ethnicity independently influence total SoE, our findings suggest that participants may specifically perceive a transformation in their real body only when both factors are matched. 

The findings from our study highlight the importance of matching both a user's ethnicity and gender to their avatar for a heightened Sense of Embodiment (SoE). Consequently, we recommend that developers and researchers prioritize efforts to align self-avatars with participants' characteristics, if achieving a robust SoE is a crucial aspect of their results, irrespective of the participants' individual characteristics.

\subsection{Facilitating Future Motion Research}
Pfeuffer et al. \cite{pfeuffer2019} argue that the analysis of VR motion data can aid in developing innovative adaptive and secure user interfaces. To facilitate diverse future investigations into the effects of participants demographics on VR experiences, we have openly shared the dataset from our study. This dataset encompasses participant demographics, such as ethnicity, age, weight, height, and VR experience. For each participant, we provide the position and rotation data for the headset, left controller, right controller, left foot tracker, right foot tracker, and hip tracker. This type of data can be leveraged for a variety of purposes, such as authenticating and identifying VR users \cite{pfeuffer2019, Miller_CombiningDataset_2022, TTI_2020,Moore_ObfuscationData_2021,Moore2023}, predicting cybersickness \cite{Islam2020,Islam2021,Jin2018}, and predicting cognitive outcomes \cite{Reinhardt2019, Moore_VelocityBasedTracking_2020, Moore_PredictingLearning_2021}.

\subsection{Limitations and Future Work}
While we attempted to recruit diverse participants, we acknowledge the limitations of our study population. The majority of our participants were young adults from the United States, and it is possible that our findings may not generalize to other populations worldwide. This is particularly relevant if a sense of racial-ethnic identity influences embodiment as we hypothesized, as these identities may vary regionally. Furthermore, our investigation only examined SoE within the context of embodied avatars.

Our results highlight the potential impact of user ethnicity on SoE and avatar matching effects in VR. To ensure spatial computing experiences are both equitable and enjoyable for diverse users, it is imperative to broaden participant demographics and explore the nuanced aspects of user characteristics. Subsequent research endeavors should focus on identifying potential explanations for the observed differences in SoE and extend investigations to diverse applications featuring embodied avatars. 

Our study represents an initial exploration of the potential impact of racial demographics on Sense of Embodiment (SoE), and we did not explore the spectrum of similarity and dissimilarity between avatars and users and self-identification \cite{fiedler_embodiment_2023}, leaving this as a promising avenue for future exploration. We specifically investigated the effects of "not matched" ethnicity and gender, yet potential variations in the degree of mismatching remain unexplored. Future work could examine participant perceptions when embodying mixed-race avatars that share only some features with the user (e.g., skin tone), contributing to a more comprehensive understanding of avatar representation in VR.

Additionally, our study solely employed an inverse kinematics-based avatar with three additional tracking points, mirroring a prevalent consumer setup for body tracking \cite{krell2023corporeal}. While this configuration offers ecologically valid insights, it is crucial to acknowledge potential variations in results when employing more advanced and accurate body tracking technologies. Future research employing sophisticated tracking systems could provide a more nuanced understanding of the relationship between avatar embodiment and user characteristics in VR. Additionally, participant demographics may also affect perceptions of IK-generated movements, which may be an avenue for future research.

\section{Conclusion}

Our study provides insight on the interaction between user demographics and avatar matching effects. By recruiting a substantial and diverse participant group, our research extends existing work, shedding light on how individual factors such as user ethnicity and gender interact with matching avatar characteristics and total SoE. Notably, we found that user ethnicity significantly affects SoE and also interacts with avatar matching effects, emphasizing the importance of considering diverse user demographics in VR studies. For example, our study found that when avatars are matched to participants' ethnicities, participants had similar Ownership. However, when avatars were not matched, Asian and Black participants were more negatively impacted. 

To ensure equitable and enjoyable spatial computing experiences for all users, future research should continue diversifying participant demographics and exploring the impact of user characteristics on VR experiences. As the VR userbase continues to evolve, understanding these dynamics is crucial for designing inclusive and immersive experiences.

\section*{Supplemental Materials}
\label{sec:supplemental_materials}All supplemental materials are available on OSF at \url{https://doi.org/10.17605/OSF.IO/YDW48}, released under a CC BY 4.0 license.
In particular, they include (1) full paper with appendices, (2) motion data.

\acknowledgments{%
	This work was supported in part by the Doctoral Research Support Award from the College of Graduate Studies at The University of Central Florida.%
}

\bibliographystyle{abbrv-doi-hyperref}

\bibliography{template}

\end{document}